\documentclass[twocolumn,showpacs,preprintnumbers,amsmath,amssymb]{revtex4}
\usepackage{graphicx}% Include figure files
\usepackage{dcolumn}% Align table columns on decimal point
\usepackage{bm}% bold math
\usepackage{amssymb}

\newcommand\mi{M\to invisible}
\newcommand\nut{\nu_e \bar{\nu}_\mu}
\newcommand\msm{\mu^+ e^- \to \nut}

\def\address{\@ifstar{\address@star}%
  {\@ifnextchar[{\address@optarg}{\address@noptarg}}}

\begin{document}

\author{S.N.~Gninenko$^{1}$}
\author{N.V.~Krasnikov$^{1}$}
\author{V.A.~Matveev$^{1,2}$}

\affiliation{$^{1}$ Institute for Nuclear Research of the Russian Academy of Sciences, 117312 Moscow, Russia \\
$^{2}$ Joint Institute for Nuclear Research, 141980 Dubna, Russia}

%\preprint{APS/123-QED}

%\title{Manuscript Title:\\with Forced Linebreak}% Force line breaks with \\
\title{ Invisible decay of muonium:\\ Tests of the standard model and searches  for new physics}

\date{\today}% It is always \today, today,

\begin{abstract}
In the Standard Model there are several canonical examples of pure leptonic processes 
involving the muon, the electron and the corresponding neutrinos which are connected by the crossing 
symmetry: i) the decay of muon 
$\mu \to e \nu_\mu \nu_e $,  ii) the inverse muon decay   $\nu_\mu  e \to \mu \nu_e $, and
iii) the annihilation of a muon and an electron into two neutrinos, $ \mu e \to \nu_\mu \nu_e $.
Although the first two reactions have been observed and measured since long ago, 
the third process, resulting in the invisible final state,  has never been  experimentally tested.
It may go either directly, or, at low energies,  via  the annihilation of a muon and an electron 
from an atomic bound state, called muonium ($M=\mu^+e^-$). 
The $M\to \nut$ decay  is expected to be a very rare process,  with the  branching fraction predicted to be $Br(M\to \nu_\mu\nu_e) = 6.6 \times 10^{-12}$ with respect to the ordinary muon decay rate. 
%The invisible decay of muonium atom $\mi$, $M=\mu^+ e^-$ 
%bound state, is expected to be a very rare process which has never been  experimentally tested.
% In the standard model it goes via  the $\mu^+ e^-$ annihilation into two neutrinos, with the  branching fraction predicted to be $Br(M\to \nut) = 6.6 \times 10^{-12}$ with respect to the ordinary muon decay rate.
Using the  reported experimental results on precision measurements of the positive muon lifetime  by the MuLan Collaboration,   we set the  first limit   $Br(M \to invisible) < 5.7 \times 10^{-6}$ (90\% C.L.), while  still leaving a big gap  of about six orders of magnitude between this bound and the predictions.  
To improve substantially the limit, we  proposed to perform an experiment dedicated to 
the sensitive search for the $\mi$ decay. A feasibility study of the experimental setup shows 
that the sensitivity  of the  search for this decay mode in branching fraction $Br(M\to invisible)$
at the level of  $10^{-12}$  could be achieved.
If the proposed search results in a substantially higher branching fraction  than predicted, say $Br(M \to invisible) \simeq 10^{-10}$ , 
this would unambiguously  indicate the presence of new physics. 
We point out that such a possibility may  occur due the muonium transition into a hidden sector and consider, as an 
example,  muonium-mirror muonium conversion in the mirror matter  model.   A result in agreement with the Standard Model
prediction would be a theoretically clean check of the pure leptonic bound state 
annihilation through the charged current weak interactions, which place constraints for further attempts beyond the Standard Model.
We believe our work gives strong motivations to perform the proposed  experiment on  search for 
the invisible decay of muonium in the near future.

\end{abstract}
%Use showkeys class option if keyword
\maketitle

\section{Introduction}

Experimental studies of particles  invisible decays, i.e.  transitions to 
an experimentally unobservable final state, played an important role both in development of the
Standard Model (SM) and in constraining its extensions \cite{pdg}. 
It is worth it to remember the determination of the number of lepton families in the SM through 
 the precision measurements of the  $Z \to invisible$ decay rate. 
In recent years, searches for invisible particle decays have  received considerable attention. 
One could mention experiments looking for extra dimensions with invisible decay of positronium (Ps=$e^+e^-$ bound state) \cite{gkr,bader},   baryonic number violation with  nucleon disappearance at SNO \cite{sno}, BOREXINO \cite{borexino}, and  KamLAND \cite{kamland}, see also Ref.\cite{tretyak}, electric charge
nonconserving electron decays $e^- \to invisible$  \cite{klap},   neutron-mirror neutron oscillations at PSI \cite{psinn} and  the ILL reactor \cite{ser}, neutron disappearance into another brane world \cite{sar}, and motivated by  various models of physics beyond the SM searches for 
invisible decays of $\pi^0$ mesons at  E949  \cite{pi0}, $\eta$ and $\eta'$ mesons
 at BES \cite{eta},
  heavy $B$ meson decays at Belle \cite{belle}, BaBAR \cite{babar}, BES \cite{bes} and
 invisible decays of the Upsilon(1S) resonance at CLEO \cite{cleo}. 
 There are also proposals for new   experiments to  search for electric charge nonconservation
in the muon decay  $ \mu^+ \to invisible$  \cite{sngmu},  and mirror-type dark matter through the 
invisible decays of orthopositronium in vacuum \cite{paolo}.

In the standard model there are several canonical examples of pure leptonic processes 
involving the muon, the electron and the corresponding neutrinos which are connected by the crossing 
symmetry: i) the decay of muon 
$\mu \to e \nu_\mu \nu_e $,  ii) the inverse muon decay   $\nu_\mu  e \to \mu \nu_e $, and
iii) the annihilation of a muon and an electron into two neutrinos, $ \mu e \to \nu_\mu \nu_e $.
Although the first two reactions have been observed and measured since long ago, 
the third process, resulting in the invisible final state,  has never been  experimentally tested.
It may go either directly, or, at low energies,  via  the annihilation of a muon and an electron 
from an atomic bound state, called muonium ($M=\mu^+e^-$). 
%The $M\to \nu_\mu \nu_e $ decay  is expected to be a very rare process,  with the  branching fraction predicted to be $Br(M\to \nu_\mu\nu_e) = 6.6 \times 10^{-12}$ with respect to the ordinary muon decay rate. 

Muonium  is a particularly
interesting system for  high
precision tests of the SM and searches for new physics. 
%Similar to Ps, it is  bounded  by the electromagnetic (e-m) 
%interaction, however contrary to Ps, which self-annihilates through the same e-m interaction,
%muonium decays into neutrino pair, $\mu^+ e^-\to \nu_\mu \nu_e$, due to the weak interactions.
Many  interesting experiments performed or planned with muonium were motivated
by tests of bound state QED in measurements of the
muonium   hyperfine splitting  \cite{qed1} and  1s-2s  interval \cite{qed2},
searches for the  lepton number violation  
 in muonium to antimuonium conversion  \cite{muconv},  tests of fundamental symmetries, 
such as CPT \cite{cpt},   probe of antimatter gravity in the free gravitational
fall of muonium \cite{gravity}, and other areas of research  \cite{vhu,kj1,jung,lauss}. 

As far as the muonium invisible decay is concerned, there  are several interesting  motivations for the 
experiment searching for the decay $\mi$ to be performed. First, the decay is predicted  to exist in the SM  at the 
experimentally achievable  today level. 
Hence, the observation of the process $\msm$ for the first time would be an interesting  test of the SM. 
Second, the decay may occur in some 
low-mass dark matter  scenarios, most of which  require coupling between the SM and hidden sectors. For instance, 
we show that  in the mirror matter model such coupling could significantly enhance 
the  $\mi$ decay rate, thus making it very attractive for direct high sensitivity searches. 
If the $\mi$ decay is observed at a rate higher than the SM prediction, it would be 
 a strong evidence for the existence of new physics  beyond the SM.

In this  paper we obtain the first limit on the decay $\mi$ and  show that 
it could be significantly improved in a  new proposed high-statistics and low-background experiment.  
 We also show that the expected level of the experimental sensitivity allows for the observation of this decay 
mode for the first time  at a rate predicted by the SM. 
 The rest of the paper is organized as follows. In Sec. II we briefly review the standard model decay rate of 
muonium and phenomenology of the muonium to antimuonium conversion. In Sec. III 
the exact mirror model,  the effect of oscillation of ordinary muonium  to the mirror one,
 and its experimental consequences are discussed.  
 The first  limit  of the branching fraction for the decay $\mi$ is obtained from available experimental data
  in Sec. IV. The experimental technique and the preliminary design of the experiment,
 detector components, simulations of the signal and background sources, as well as the expected sensitivity  are discussed  in detail in Sec.V. Section VI contains concluding remarks.

\section{Muonium decay in the Standard Model  and  phenomenology of $M\to \bar{M}$ conversion}

The muonium atom consits of a positive muon and an electron, which are leptons from 
two different generations. To our current knowledge these are particles without any known internal structure. This makes muonium  an ideal system for testing QED and  fundamental symmetries in physics, and  allows us to calculate 
muonium properties  to very high accuracy within the framework of the bound state QED. 
For example, for the hyperfine structure the theoretical predictions and 
measurements agree  substantially better than for hydrogen atoms \cite{karsh}. 

As discussed previuosly,  muonium atom, similar to the lightest known exotic hydrogen-like 
atom positronium,  is bounded by the electromagnetic (e-m)  interaction. 
However, differently from positronium the muonium cannot  self-annihilate through the 
 e-m interaction because it would violate the lepton number conservation. 
Instead, the SM allows  the self-annihilation of muonium  into neutrino pair through the 
lepton number conserving weak interaction.  
At the current level of experimental
and theoretical precision these are  the only interactions present in the muonium system .

In the SM  muonium is unstable mainly due to the decay of muon 
$\mu^{+} \rightarrow e^{+}\nut$. Its decay rate 
in vacuum coincides with those of the muon decay given by 
\begin{equation}
\Gamma(M\to e^{+}\nut  e^-) =
\Gamma(\mu^{+}  \rightarrow e^{+}\nut) =
\frac{G^2_{F}m_{\mu}^5}{192\pi^3}\,.
\label{muord}
\end{equation}
Here $G_{F} = 10^{-5}m_{p}^{-2}$ is the Fermi constant and $m_{p}, ~m_\mu$ are  
the proton and muon masses, respectively. In matter, muonium  is
typically formed in the singlet or triplet state,  with the total angular momentum equal 0 or 1, respectively.
The SM  predicts  direct annihilation 
of the triplet muonium, $J^{PC} = 1^{--}$ bound state, into neutrino antineutrino pair
$M \rightarrow \nut$ with a very small decay rate. 
The corresponding branching fraction $Br(M\to \nut)$ is calculated  to be \cite{li}
\begin{eqnarray}
Br(M\to \nut)=\frac{\Gamma(M \rightarrow  \nut)}{\Gamma(\mu^+ \rightarrow e^+\nut)} =  \nonumber \\
=48\pi\alpha^3 (\frac{m_{e}}{m_{\mu}})^3  \approx 6.6\times 10^{-12},
\end{eqnarray}
where $\alpha$ is the fine-structure constant, and $m_e$ is the electron mass. This result was further confirmed in \cite{marci}. The singlet muonium cannot decay into two (massless) neutrinos, as it contradicts to momentum and angular  momentum conservation simultaneously. 

Very interesting feature of muonium  is 
the possibility of conversion(or oscillation) into its  
antiatom, i.e. the $\mu^- e^+$ bound state \cite{Pontecorvo,wf1,wf2}. This  reaction 
violates the conservation of lepton flavor numbers by two units 
($\Delta L_{e,\mu} = \pm 2$)  that makes searches for the $M - \bar{M}$ 
conversion especially interesting due to experimental observations of lepton flavor violation in  neutrino 
oscillations. 

The existence of muonium to antimuonium conversion is predicted in  different extensions of 
the SM.  
The simplest way to understand  the phenomenology of $M - \bar{M}$ conversion is the use of the effective four fermion interaction of the 
$(V - A)(V - A)$ type \cite{wf1}, namely 
\begin{equation}
L_{M\bar{M}} = (\frac{G_{M\bar{M}}}{\sqrt{2}})\bar{\mu}\gamma_{\lambda}
(1 - \gamma_5)e \bar{\mu}\gamma^{\lambda}(1 - \gamma_5)e + H.c.,
\label{muc}
\end{equation}    
where $G_{M\bar{M}}$ is a coupling constant characterizing the strength of a new 
flavor violating interaction. In the absence of an external magnetic 
field the muonium and antimuonium have the same ground-state energy levels.  
Flavor violating interaction (3) would cause a splitting of their energy 
levels by amount \cite{wf1,wf2} 
\begin{equation}
\delta \equiv 2 <\bar{M}|L_{M\bar{M}}|M> = \frac{8 G_F}
{\sqrt{2}n^2 \pi a^3_0}(\frac{G_{M\bar{M}}}{G_F})\,.
\label{del}
\end{equation}
Here $n$ is the principal quantum number of the muonium atom and $a_0 = 
\frac{m_e + m_{\mu}}{m_e m_{\mu} \alpha}$ is the Bohr radius of the 
muonium. 
Numerically, for the ground state of muonium $(n = 1)$  
\begin{equation}
\delta = 1.5 \times 10^{-12}(\frac{G_{M\bar{M}}}{G_F}) (eV)\,.
\end{equation}

The $M - \bar{M}$ conversion is analogous to the $K^0 - \bar{K}^0$ mixing. 
If a muonium atom is formed at $t = 0$ in  vacuum,  it could oscillate 
into an antimuoniumm atom. For a small $t$ value the probability of the 
oscillation is represented in the form \cite{wf1}
\begin{equation}
P_{M\bar{M}}(t) = \sin^2(\frac{\delta t}{2})\cdot \Gamma_{\mu}
e^{-\Gamma_{\mu}t} \approx (\frac{\delta t}{2})^2\cdot \Gamma_{\mu}
e^{-\Gamma_{\mu}t}\,.
\end{equation}  
Here  $\Gamma_{\mu} \equiv 
 \Gamma(\mu^{+} \rightarrow e^+ \nut)$ 
is the muon decay width.
The total conversion probability after integration over time is equal to
\begin{eqnarray}
P_{M\bar{M}} = \int_{0}^{\infty}\rho_{M \bar{M}}(t) dt = 
\frac{|\delta|^2}{2(|\delta|^2 + \Gamma_{\mu}^2)}  \nonumber \\
= 2.56 \times 10^{-5}\cdot (\frac{G_{M \bar{M}}}{G_F})^2.
\end{eqnarray}
The best current experimental limit on the $M\to \bar{M}$ conversion leads 
to bound $|G_{M \bar{M}}| \leq 0.003 \cdot G_{F}$ 
\cite{muconv}. 
It should be noted that in the presence of external electromagnetic fields or  collisions with  residual gas
molecules the $M - \bar{M}$ transitions become suppressed. 

\section{Muonium conversion in mirror matter model}

The idea that along with the ordinary matter may exist its exact mirror copy is an old 
one \cite{ly,kop}. This  new hidden gauge sector is predicted to exist if parity is the  unbroken symmetry  of nature, for an excellent review see  Ref. \cite{okun}. In accordance with this idea each ordinary particle of the SM has a  corresponding mirror partner of exactly the same mass as the ordinary one. 
In the modern language of gauge theories, the mirror particles are all singlets under	
the	standard	 $G=SU(3)\otimes SU(2)_L\otimes U(1)_Y$ gauge interactions \cite{flv, zurab, mirror}. Hence, 
they couple to the ordinary particles either by gravity or by other very weak forces.
In this model the parity is conserved because	the	mirror	particles have the right-handed (V+A)  mirror weak interactions while the ordinary	particles	experience	the	usual left-handed weak interactions. Mirror matter is dark in terms of the SM interactions,  and could  be a good candidate for dark matter, see,  e.g.,  Refs.\cite{mirror,zbcos,ciar,ciar1,zurab1,zbast}. For instance, 
it is argued that  annual modulations of the signal observed by the DAMA Collaboration are caused by the mirror dark matter scattering in their detector \cite{footdm1,footdm2,footdm3,footdm4}.  
 
The gauge group of our world and mirror 
world is assumed to be \cite{flv}
\begin{equation}
G_{tot} = G_{SM} \otimes G_{M},
\end{equation}
where the SM gauge group $ G_{SM}$ coincides with the mirror world gauge group $G_{M}$.
The interaction between our and mirror sectors could be transmitted by some gauge singlet particles
interacting with both sectors. Such kind of interaction could explain the baryon asymmetry of the Universe \cite{zbnn1},
some fraction of dark matter in the Universe \cite{zbast}, and  can also results in the particle mixing and oscillation between the ordinary and mirror sectors. Any  neutral, elementary or composite  particle, in principle,   can have mixing with its mirror duplicate, such as   photon-mirror photon \cite{holdom,glash}, neutrino-mirror neutrino \cite{flv,zurab}, etc. which can be experimentally tested. 
   For example, the neutron-mirror neutron mixing  via a small mass term $\epsilon (nn' + n'n)$ proposed in \cite{zbnn1}, results in ordinary    neutron anomalous disappearance, in addition to their decays or  absorption
   due to SM interactions. At present, there are performed and proposed searches 
 for mirror matter via  the invisible decay of orthopositronium in vacuum 
 \cite{paolo,holdom,glash,sng,fsng,gkmr,opsjap,sngbeam}, through neutron-mirror neutron oscillations 
 \cite{zbnn1,zbnn2,zbnn3}, and via Higgs- mirror Higgs mixing  at LHC \cite{igv,ch1,ch2}.

\begin{figure}
\includegraphics[width=0.45\textwidth]{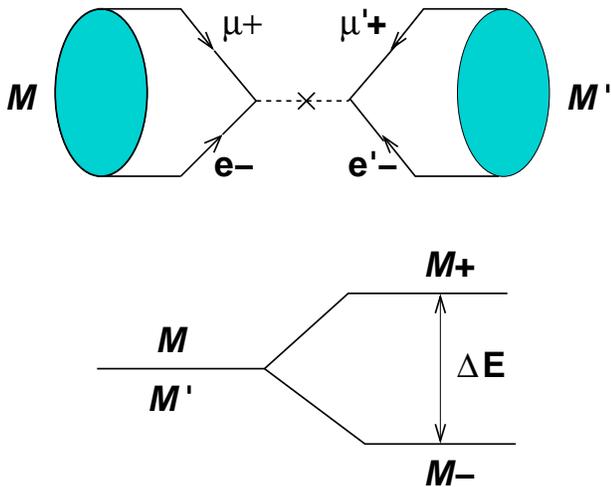}% Here is how to import EPS art
\caption{\label{fig:setup} 
The double degeneracy between mass eigenstates of ordinary ($M$) and mirror 
($M'$) muonium is broken when a small mixing  is included.}
\label{muoni}
\end{figure}

 As  mentioned previously, in a mirror world model, there must exist the mirror  muonium which is the bound state of a mirror muon $\mu '$ and a  mirror positron $e'^+$ with the same mass and decay width as the ordinary muonium.   
We could assume  the existence of a (super)weak interaction invariant under 
the gauge group $G_{tot}$, which allows transitions between the ordinary and mirror muonium. Such effective interaction
 can be written in a form analogous to \eqref{muc} 
\begin{equation}
L_{M M'} = (\frac{G_{M M'}}{\sqrt{2}})\bar{\mu}\gamma_{\lambda}(1 + \gamma_5)e 
\bar{e}'\gamma_{\lambda}(1 - \gamma_5)\mu' + H.c.
\label{mirr}
\end{equation}  
where $G_{M M'}$ is a coupling constant characterizing the strength of the $M-M'$ transition, and 
 $e'$ and $\mu'$ are the mirror electron and muon fields, correspondingly.
 The interaction \eqref{mirr} leads to conversion of ordinary  muonium to mirror muonium, as schematically illustrated in Fig. \ref{muoni}.
  The interaction \eqref{mirr} breaks the degeneracy between $M$ and $M'$ states, 
 so that the vacuum energy eigenstates are $M+=(M + M')/\sqrt{2}$ and $M-=(M - M')/\sqrt{2}$, 
 which are split in energy by $\Delta E$ given by an expression similar to \eqref{del},  see Fig.\ref{muoni}. 
 The interaction eigenstates are maximal combinations of mass eigenstates which implies 
that $M$ oscillates into $M'$. Thus,
 a system which is pure muonium at $t=0$ will develop an admixture of mirror 
muonium at a later time.  In closed analogy with the case of muonium antimuoniumm conversion one can find that the probability of seeing the system in vacuum decay as mirror muonium
$M'\to e'^+ \nu'_e \nu'_\mu + e'^-$ rather than as ordinary muonium $M\to e^+ \nut+ e^-$
is given by 
\begin{equation} 
P(M')= \frac{1}{2}\frac{\delta^2}{\delta^2+\Delta^2+\lambda^2}
\end{equation} 
where $\lambda=0.45\times 10^6~\rm{sec^{-1}}$, or $3\times 10^{-10}$ eV, and $\Delta$ is any additional 
splitting of $M$ and $M'$, by external electromagnetic fields.  A detailed discussions of the effects of collisions and
external fields on oscillation probability of a similar system, positronium, can be found in Ref.\cite{dgt}. Estimating $\delta$ by using (9)
results in the integral probability 
of muonium to mirror muonium conversion determined by 
(for $\Delta = 0$)
\begin{equation}
P_{M M'} = 
2.56 \times 10^{-5}\cdot (\frac{G_{M M'}}{G_F})^2  \,.
\label{mir}
\end{equation}
Because mirror muonium decays into a mirror electron, positron, and neutrinos  the 
experimental signature of 
the  $M-M'$ conversion is the invisible decay $\mi$ of the ordinary muonium in vacuum. 
Current bounds on coupling constant 
$G_{M M'}$ are rather weak. For $G_{M M'} = G_{F}$ the branching fraction 
of the ordinary muonium decay into invisible mirror state is
$Br(M \rightarrow invisible) = 2.56 \times 10^{-5}$, which is seven orders of magnitude 
higher than  those  predicted by the SM branching fraction of Eq.(2). Thus, 
we see that in the mirror model it is possible to have nonzero conversion 
of our muonium to mirror  muonium. The signature of such conversion is the  decay $\mi$ 
 and, moreover, it is possible to expect some enhancement of this decay rate.
  Similar to the $M-\overline{M}$ conversion,  the probability $P_{MM'}$ can 
be affected by an additional splitting of $M$ and 
$M'$ states due to  an external electric or magnetic field \cite{wf1,dgt}. 
It might also be suppressed, if there is a high collision rate of muonium atoms with the cavity walls or residual gas 
molecules in the experiment.

Note that some extensions of the SM allow the $\mi$ decay. For instance, 
in the model with the additional sterile neutrino, interaction
$$ L = G'\bar{\mu}\gamma_{\nu}(1 - \gamma_{5})e \bar{\nu}_s\gamma^{\nu}(1 -\gamma_{5})\nu_{s} +h.c. $$
results in  invisible muonium decays into sterile neutrino $  M \rightarrow \nu_{s}\bar{\nu}_{s} $. However, 
constraints obtained from the agreement between the measured and predicted properties of the
$\mu$ decay \cite{pdg} lead to a strong bound $\Gamma(M \rightarrow \bar{\nu}_s \nu_{s} \leq O(10^{-2})
\Gamma(M \rightarrow \nu_{e}\bar{\nu}_{\mu})$ on sterile neutrino decay width, which makes it not 
very exciting for further consideration.

\section{Indirect experimental limit on the $\mi$ decay}
Consider now bound on the invisible decay of the $M$ state, which can be obtained from existing experimental data.
If an exotic  $\mi$ decay  exists, it would contribute to the total muonium decay rate:
\begin{equation}
\tau_{M}^{-1}= \Gamma_{M} (M \to all) = \Gamma_\mu + \Gamma(\mi)+...
\label{rate}
\end{equation} 
and, hence decrease the determined muonium lifetime  $\tau_M$.

In order to estimate the allowed extra contribution of $\Gamma(M \to invisible)$ to Eq.(\ref{rate}), and 
to obtain the limit on the branching fraction $Br(M\to invisible)$,  
we use the results on precision measurements of the positive muon lifetime reported by 
the MuLan collaboration \cite{mulan1,mulan2,mulan3}. In these measurements two different  
targets were used in the detector to stop muons. For the first one, the magnetized ferromagnetic alloy target 
(AK-3), the dominant population was stopped $\mu^+$'s, and the corresponding muon decay lifetime was
measured to be \cite{mulan1} 
\begin{equation}
\tau_\mu^{AK-3} = 2.1969799\pm 0.0000027 ~\mu{\rm s}
\label{tauak}
\end{equation}
For the second one, the quartz (SiO$_2$) 
target, the dominant species were muonium atoms formed by stopping muons 90\% of the time, and  
the muon lifetime was found to be  \cite{mulan2}
\begin{equation}
\tau_\mu^{Q} = 2.1969812\pm 0.0000038 ~\mu{\rm s}
\label{tauq}
\end{equation}
It is important to note, that in the framework of the SM the possible lifetime difference between the muonium 
atom in the quartz target and free muon in vacuum was estimated to be negligible,  of the order of 1 part per billion \cite{marci}.  
By comparing  the measured  muonium decay rates from Eq.(\ref{tauak}) and Eq.(\ref{tauq}),
and assuming that the fraction of triplet muonium state  in the quartz 
target  is 3/4, one finds that the upper limit on the branching fraction of the decay $\mi$ is    
\begin{equation}
Br(M \to invisible) < 5.7 \times10^{-6}
\label{muon}
\end{equation}
at the  90 \% C.L..
There are still six orders of magnitudes difference between 
the limit of Eq.(\ref{muon})  and the SM prediction (2). Note that the obtained result cannot be used to constrain
$M-M'$  oscillations in vacuum because of their high collisional suppression in the quartz target.  
In the next section we show how the limit of Eq.(\ref{muon}) can  be significantly   
improved  in the new proposed experiment.

%%%%%%%%%%%%%%%%%%%%%%%%%%%%%%%%%%%%%%%%%%%%%%%%%%%%%%%%%%%%%
%%%%%%%%%%%%%%%%%%%%%%%%%%%%%%%%%%%%%%%%%%%%%%%%%%%%%%%%%%%%%

\section{Direct experimental search for the  $M\to invisible$ decay}

The decays $\mi$  are rare events and their observation presents a challenge for the detector design and performance. 
Here, we focus mainly on discussions of the experimental setup to search for the decay $\mi$ in vacuum, which is 
also sensitive to the muonium -  mirror muonium conversion. The similar setup without vacuum requirements
is simpler, it would provide  better sensitivity and can be used for the first observation of the decay $\mi$ and 
search for new exotic channels  of this decay mode which are not affected by the presence of matter of external fields.   

The main components of the experimental setup to search  for the invisible 
decay of  muonium  are schematically illustrated in Fig. \ref{fig:setup},  see also \cite{sngmu}.
The setup is equipped with a high efficiency muon tagging system, high hermeticity electromagnetic 
calorimeter (ECAL), and an intelligent  trigger system. 
The experiment employs a surface $\mu^+$ beam, which is produced in a target and 
transported to the detector in an evacuated beam line tuned to $\sim 26~ \rm{MeV/c}$.
Such the world's  brightest continuous surface muon beam with intensity $\simeq 10^7~\mu$/s is  available 
 at the Paul Scherrer Institut (PSI) \cite{abela}. This  beam  was used, for example 
for a sensitive search  for $M-\overline{M}$ conversion \cite{muconv}.
Positively charged muons pass through  $\sim 100~\mu$m thick beam counters (S$_{1,2}$)
are focused into a vacuum cavity through a narrow aperture 
closed by the beam counter S$_3$, and, after passing through the counter S$_4$,  
strike the SiO$_2$ aerogel (or SiO$_2$ powder)  target  ($T$) used for the muonium atom formation \cite{mutem,sio2}.
The energy of entering muons is degraded by the counters material  to maximize 
the muon stopping rate. 
Muonium atoms are formed by the electron capture with efficiency $\simeq$60 \% per $\mu^+$ 
stopped in the target.
Most of the atoms emerge from the target grains into the intergranular
voids. With a mean-free path of $\simeq$ 1 cm, muonium 
is able to diffuse through the network of voids over distances
 longer than the target  thickness and escape through
the surface into vacuum \cite{will}.  Muoniums undergo collisions of the order
$10^5 - 10^7$  with the silica grain walls, those number depends on  the depth of muonium formation, and   approach thermal equilibrium.  Then, on average, 3.3\% of them leave the
target surface with thermal Maxwell-Boltzmann velocity
distribution at the  temperature of the target \cite{mutem}. The fact that 
muonium confined to the voids is expected to be almost  fully thermalized,  
was  confirmed by a separate experiment on $M$'s emitted into vacuum from 
a mesaporous silica film at cryogenic temperatures \cite{anto}. Although 
the $M$ kinetic energy distribution
is nominally that of Maxwell-Boltzman emission, one might expect a higher-energy tail of $M$'s formed from backscattered
muons that never approached thermalization. These $M$ events are distinct from the thermal $M$'s
 that have diffused out of the target; however, their intensity is expected to be very small.
 The fraction of muonium atoms produced in the target that  decay either in the target or in vacuum 
 can be determined relative to the number of muons on the target with a technique described in Ref.\cite{muconv}.
 
\begin{figure*}
\includegraphics[width=0.9\textwidth]{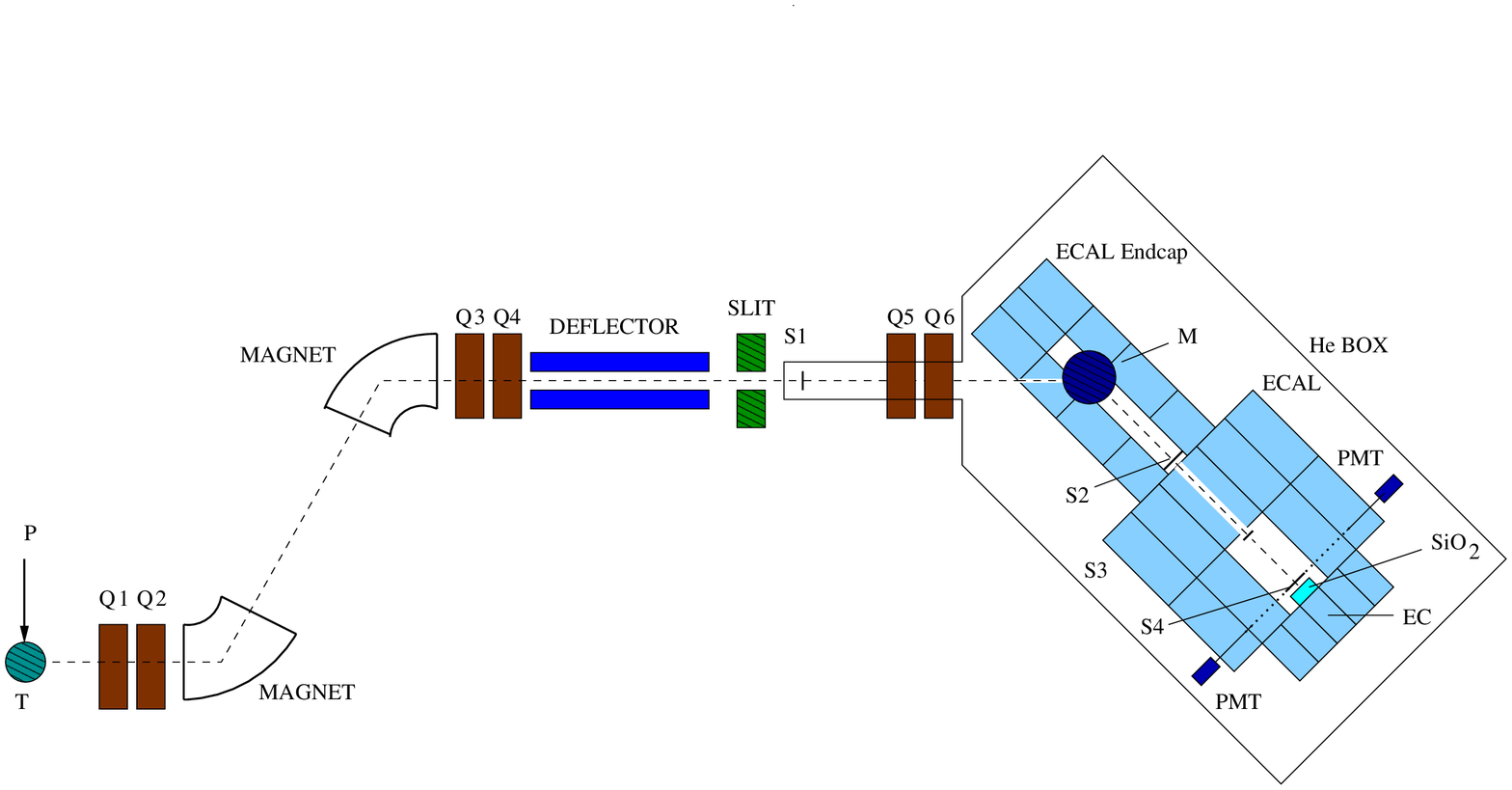}% Here is how to import EPS art
\caption{\label{fig:setup} Schematic illustration of the experimental setup 
to search for  the muonium invisible decay. The beam of surface $\mu^+$'s passing through 
the beam defining counters S$_{1-4}$
is focused by quadrupole magnets (Q5,Q6) into a vacuum cavity through a narrow aperture and  strikes the SiO$_2$ aerogel target  ($T$) used for the muonium atom formation. 
Shown are also the 4$\pi$ hermetic  BGO electromagnetic calorimeter (ECAL), the 
ECAL endcap counter (EC) used as a light guide for the light produced in the beam counter S$_4$, 
and the magnet (M)  used to deflect the beam.
The counters S$_{1-3}$ and the upstream ECAL counters are also used as a veto against photons, decay positrons or backscattered muons that could escape the cavity through the entrance aperture. The deflector is used 
to operate the setup in a "muon on request" mode.}
\label{setup}
\end{figure*}

 The target is surrounded by a hermetic 4$\pi$ ECAL to detect energy deposition from the decay $M \to all $ of 
muoniums produced  in $T$.  
 As shown in Fig. \ref{fig:setup}, before muons reach the entrance to the vacuum cavity, they bend in a magnetic field.   
The purpose of employing the magnet is threefold: (i) to provide a transverse kick to positive muons in order 
to allow them to enter the vacuum cavity through the narrow aperture, (ii)
to detect photons, positrons, or muons  that could escape the cavity through the entrance aperture by a set of ECAL counters
placed around the muon bend region, and iii) to enhance identification of positive  muons entering the 
calorimeter. This additional detector is placed up stream of the entrance aperture, as shown in Fig. \ref{setup}. The deflector is used in order to operate the setup in a "muon on request mode" with the 
repetition rate in the range 200-400 kHz.

The  energy deposition readout in the ECAL  is triggered by a tag signal of the muon 
appearance on the target, which is defined as the  coincidence of the four  
 signals from a muon passing the beam counters $S_{1-4}$. 
To enhance significance of the muon tag  the  time-of-flight information can  be used.
 For  the muon beam momentum of 26 MeV/c , the latter  corresponds to about 
 1.3 ns  per 10 cm of the muon path length. 

For  the ordinary muonium decay \eqref{muord} the experimental signature is the ECAL energy deposition
 from a single decay positron with energy $E_{e^+} = m_\mu - E_{\nu_e} - E_{\nu_\mu}$, where $  E_{\nu_e},  E_{\nu_\mu}$ are the electron and muon neutrino energy, respectively.
The experimental signature of  the $\mi$ decay
is the apparent disappearance  of the energy deposition $E = m_\mu + m_e$ in the ECAL.    
In other words, the signature of the $\mi$ decays is 
   an event with  the 
 sum of the ECAL crystal energies deposited by  the final-state particles equal to zero. 
Zero energy is defined  in this case as an
energy deposition  below a certain ECAL energy threshold,  $E_{tot} < E_{th}$.   
%peak from mono-energetic positrons in the low  energy part of the 
%positron spectrum corresponding to the principal mode of $\mu ^{+}$ decays.\
The expected distribution of energy deposited in the  ECAL 
 from  $\mu^+$'s stopped in the target is shown in Fig. \ref{spect}. The distribution is 
  a  sum of two spectra from $\mu^+ \to all$ and 
  $M \to all$ decays  and is discussed in detail below in Sec.A.

To estimate the sensitivity of the proposed experiment 
 a feasibility study  based on GEANT4 \cite{geant}
Monte Carlo simulations have been  performed. The beam of positive muons is 
stopped  in the central part of the cylindrical  target of SiO$_2$ aerogel  
 with the density about 35 mg/cm$^3$ and
of thickness 8 mg/cm$^2$ and is supported in vacuum by an aluminum foil with an 
inclination with respect to the muon beam axis. A similar target  was 
previously used as a convertor of muon to muonium atoms in the experiment of Ref.\cite{muconv}.
       The  ECAL is an array of $\simeq 100$  BGO counters each of 
52 mm in diameter and  220 mm long, which was  previously used in the PSI experiment on 
precise measurements of the $\pi \to e+\nu $ decay rate \cite{pienu}. Timing and energy deposition information from 
each BGO crystal can be digitized for each event.
The processing of the BGO counter signals is described in detail in Ref.\cite{pienu}, see also Refs.\cite{bader,pcth}.

\subsection{Background  for the  $M\to invisible$ decay}
 
The background processes for the $\mi$ decay can be classified as being due to 
beam-related, physical, and detector-related backgrounds.
 To investigate these backgrounds down to the level  $Br(\mu \to invisible) \lesssim 10^{-12}$ with the full 
 detector simulation would require 
the generation of a very large number of muon decays resulting in a
prohibitively large amount of computer time. Consequently, only the most 
dangerous background processes are considered  
and estimated  with a  smaller statistics combined  with numerical calculations.
\begin{figure}
\includegraphics[width=0.55\textwidth]{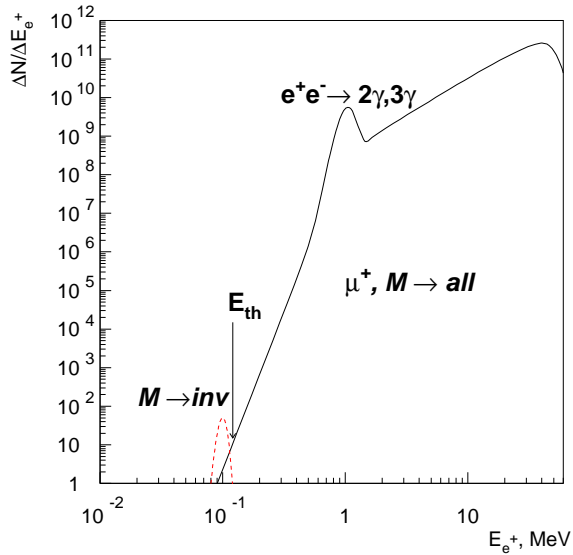}% Here is how to import EPS art
\caption{\label{decay} The expected distribution of energy deposition in the ECAL
from  $8\times 10^{12}$ muons stopped in the target, corresponding to the  
decays $\mu^+\to all$ and  $ M \to all$.   
The peak around 1 MeV corresponds to 
energy deposition from the  $e^+e^- \to 2\gamma, 3\gamma$ annihilation of decay positrons stopped in the vacuum cavity.
The arrow shows the energy threshold for the decay $\mi$ detection.
The dashed curve represents the signal from the decay $\mi$ if it 
exist at the level predicted by the SM. }
\label{spect}
\end{figure}

The  beam-related backgrounds produce the fake muon tag and  
can be  categorized as being
due to a  beam particle misidentified as a muon, or several 
 beam particles  which produce a fake muon tag due to 
accidental coincidence of signals from S$_{1-4}$.
The first type of   background occurs, e.g., due to the production
of slow protons in the target, which enter the detector and produce zero decay energy.
 Incoming neutrons   could  scatter in  the S$_{1-4}$ and being accidentally misidentified as $\mu^+$ could  
 also be contributed to the beam
background. Identification of the incoming
particle as a muon based on the  requirements of the delay by the muon time-of-flight 
coincidence between the beam counter signals 
suppresses the single-beam background down to the level $< 10^{-13}$. This estimate 
is obtained under the assumption of having Gaussian shape with the time resolution of $\simeq 1$ ns
for the distributions of time of flight between counters S$_{1-4}$. It is also assumed that the  
 admixture of the other charged particles in the  beam is 
 below 1\%, which  although depends on a particular experimental environment. 

For the design shown in Fig.\ref{setup}, the required efficiency for the $M$ decay energy detection 
can be obtained only by keeping the amount of passive material in the region of vacuum cavity 
as small as possible. For example, to remove dead materials from the vacuum cavity walls 
the cavity could be made directly in a big single crystal or out of a few ECAL central crystals. For example, the light signals 
produced in the $S_4$ scintillator counter could be  readout through the SiO$_2$ transparent target and 
the  ECAL endcap crystal  which acts as a light guide, see Fig. \ref{fig:setup}.
The $S_4$ signals could be distinguished from the endcap crystal signals due to their significantly 
different decay times by using the technique described in detail in Ref.\cite{bader}.
In the presented simulations we did not consider too complicated design of the setup and try to keep it
as realistic as possible. The reported further analysis takes into account active materials of the ECAL,
 passive materials from  the target, vacuum cavity walls,   and from   
the ECAL crystals and the target wrapping. The following main sources of
 physical- and detector-related backgrounds are identified and evaluated:

\begin{itemize} 
\item the principal muon decay $\mu^+ \to e^+ \nut$
 into the final state 
with the positron kinetic energy $E_{kin}$  less than the 
 detection energy threshold $E_{th}$ ( $\simeq$ 100 keV). Indeed, if  
 $E_{kin} < E_{th}$ the event   
 becomes invisible. 
 
 To suppress this background, one has to use  as low a threshold as possible  
and to  performed the experiment with a well separated  positive muon beam 
with an extremely small contamination of negative  
pions or muons which could mimic the true signal. However, even if the positive muon 
decays into a low energy positron that stops in the cavity,  the latter would  
annihilate into two (or three) photons at a lifetime scale of the order of a few ns.
Thus, for such events,  
 the  minimum energy deposition in the ECAL 
will be $m_{e^+} + m_{e^-} \simeq 1$ MeV, i.e. well above the 
threshold, making these  events visible; see Fig. \ref{spect}. 

Another  way to lose the decay energy 
is due to the annihilation gammas   photoabsorption and/or 
Compton scattering in the target. In this case, when  almost all annihilation energy is deposited in $T$
the event becomes invisible, which results in a fake $\mi$ signal.
To suppress this background,  the target should be optimized in size and 
made of a low-Z material to minimize the 
crosssection of the photoabsorption which is $\sigma_{pha}\sim Z^5$. For example, for 
a target made of a plastic scintillator, the probability of both 511 keV photons energy absorption 
in a volume of $\simeq1$ cm$^3$ is found to be less than 10$^{-8}$ \cite{sngmu}.
In the SiO$_2$ target with the density 35 mg/cm$^3$ the effect is smaller. 
The ECAL  efficiency with respect to detection of energy from the positron annihilation was checked in the 
 experiment  \cite{bader} on the search for the $Ps\to invisible$ decay. 
For events corresponding to $ 2,3 \gamma$ annihilation of  $e^+ e^-$ pairs at rest 
with the ECAL energy deposition $\sim 1$ MeV, the upper limit on the branching fraction of the reaction
$e^+ e^- \to invisible$ was found to be $Br(e^+ e^- \to invisible) \lesssim  10^{-8}$ at 90\% C.L. for the 
ECAL energy threshold of 80 keV.

The absorption of annihilation photons in the cavity materials has been studied in the proposal 
on the search for the $oPs\to invisible  $ decay in the vacuum of Ref. \cite{paolo}.  
Simulations show that the main contribution to the $\gamma$ 
inefficiency comes from the total (due to photoabsorption) or fractional 
(due to Compton effect) photon energy loss in the material of the vacuum cavity.
To suppress this background  the cavity should be  made of a low-Z material to minimize the  
cross section of the photoabsorption. 
Distributions of the energy deposited in the dead material surrounding the target region 
 from annihilation events in the target were 
obtained with simulations  for a 0.84 mm thick aluminum pipe and a composition  pipe made of 
 0.04 mm aluminum and 0.800 mm carbon. For the later case the fraction of simulated $2\gamma$ 
events with the  energy absorbed in the cavity walls $> 900$ keV, i.e. energy deposited  in the ECAL is $E < 100$ keV,
  was found to be $< 10^{-8}$.
  
In Fig. \ref{fraction}, the partial muon decay rate $\Delta \Gamma_\mu$ 
into a  positron with  $E_{e^+} < E_{th}$  is  shown as a function of $E_{th}$.
Taking into account that energy of positrons   that stop in the cavity is  typically   $E_{kin} < 2-3$ MeV, 
the  fraction of such $e^+$'s is estimated to be $ P_T \lesssim 10^{-4}$.
 Combined   probability   to get energy deposition in the ECAL   from positron annihilation in the cavity $< 100$ keV is
 estimated to be   $ P_{2\gamma} \lesssim  2\times 10^{-8}$.  Therefore, this  background from inefficient detection of 
 low energy positrons  allows  potentially to reach sensitivity in the branching ratio of the invisible muonium  decay  as small as
$ P_T \cdot P_{2\gamma} \simeq 2\times 10^{-12}$,
assuming  the detection energy threshold is as low as $E_{th} \simeq 100$ keV.
\begin{figure}
\includegraphics[width=0.55\textwidth]{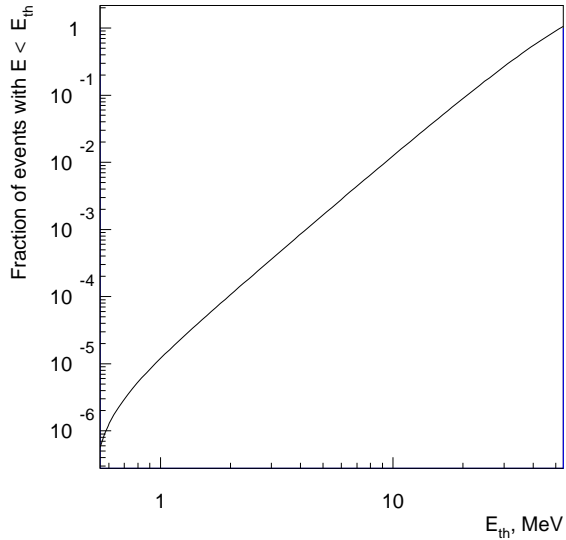}% Here is how to import EPS art
\caption{\label{decay} The fraction of events from the  decay $M\to e^+ \nut + e^-$ 
with the positron energy $E<E_{th}$ as a function of $E_{th}$. }
\label{fraction}
\end{figure}

\item 
the loss of the muon decay energy in rare processes of energetic  positrons 
$e^+ +  A \to invisible $ with an invisible final state
could be induced  either by electromagnetic or weak interactions of the positron.
For example, $e^+$ could lose almost all its energy due to emission  
of a hard photon in the bremsstrahlung process  in the target or in the ECAL. The photon could either 
penetrate the calorimeter without interactions, or could be photoabsorbed by an atomic nucleus  
resulting in the invisible final state consisted of secondary neutrons. 
However, due to the charge conservation,  there is always a 
low energy positron in the final state,
  which produces $\simeq$ 1 MeV energy through the $e^+e^-$ annihilation of the positron at rest, thus making 
the event visible. Combined analysis results in this background level $\lesssim 10^{-13}$.
The background from an energetic positron conversion into proton through the reaction 
$e^+ + n \to p + \bar{\nu}_e$ induced by the  charged current 
 weak interaction is found to be negligible. 

\item Another possible background  could be due to the excitation of a long-lived nuclear state 
via the radiationless annihilation of an energetic  positron with a K-shell 
electron. This is a 3-body reaction $e^+ + e^- + A \to A^* $, where the $e^+e^-$ annihilation energy 
is  absorbed  by the  nucleus $A$. 
The cross section for such a reaction has not yet been studied in detail for the 
wide class of nuclear isotopes and full range of positron energies. By using the available upper limit 
 on the resonant cross section $\sigma_{e^+} < 4.3\times 10^{-26}$ cm$^2$ at 99\% C.L. 
obtained for isotope $^{115}$In with   a mono-energetic positron beam of about 90 keV kinetic 
energy \cite{lynn} we estimate this background to be   $\lesssim 10^{-13}$, 
assuming the  $^{115}$In contamination in the cavity and target materials to be at the level below 1 ppm. 
More detailed study of this background source is required. Note, that in 
 principle, it is possible to excite a nucleus long-lived state with a lifetime $\tau \gtrsim~ 60 \mu$s. 
 However, such excitation  levels are present in specific isotopes, such as  $^{115}$In, 
 whose admixture is expected to be small.

\item incomplete ECAL hermiticity:
our  study   identified a possible background to the signal
as due to energetic decay positrons escaping the detection region  
 though the cavity entrance aperture. This effect  increases the 
 disappearance rate of muonium and therefore must be addressed. 
Consider, e.g. the case when  a muon decays either in flight or in the target into a fast positron 
with momentum pointing exactly to the entrance 
aperture. Then, the decay positron could be undetected in beam counters S$_3$ and S$_2$ due to their
 inefficiency. The same effect could occur if the incoming muon backscatters either in S$_4$ or in the target 
without losing too much energy, and escapes the detection in counters S$_{2,3}$.
However, due to the presence of the magnetic field in the vicinity of the entrance to the cavity,   the 
 trajectory of the escaping positron or muon is bent up and  it would be detected by the ECAL counters.

The probability for a particle to escape detection in this case can be estimated as 
\begin{equation}
P_{esc} \simeq P_a \cdot P_m  \cdot \zeta_2 \cdot \zeta_3 \cdot \zeta_{ECAL}
\end{equation}
where $P_a, \zeta_2$,    $\zeta_3$, and $\zeta_{ECAL} $  are, respectively, the probability for a particle to 
pass  through the entrance aperture, the inefficiencies 
of beam counters  S$_2$, S$_3$ and ECAL counters  to detect the particle.   

To suppress this type of  background 
the entrance aperture should be reduced in size as much as possible and 
should be  closed by as  high as possible efficiency  counters S$_{2,3}$, 
as shown in Fig. \ref{fig:setup}, which act as the beam defining and also as the veto counters.
Then, the  background could be  suppressed by requiring an absence of activity in the beam counters 
after detection of the incoming muon. 
Assuming  isotropic distribution of backscattered muons or decay positrons, inefficiency  for  particle detection in S$_{2,3}$  of $\simeq 10^{-2}-10^{-3}$, the diameter of the entrance aperture of $\simeq$1 cm, and inefficiency of the upstream ECAL detection $\simeq 10^{-4}$  leads to  the final suppression of this source of background down to the level of at least $\simeq 10^{-13}$.

\item The leak   of  muonium atoms  through the entrance aperture into the region of 
lower detection efficiency could also contribute to  the  
disappearance rate of muonium. However, assuming that muoninum leaving  the target is 
thermalized and has kinetic energy below eV (300 K), the effect is suppressed 
to a negligible level by closing the aperture with the counter S$_3$, as shown in Fig. \ref{setup}.
\end{itemize}

In Table I contributions from the previously discussed  background processes are summarized. 
The dominant background source is due to the absorption by passive materials of  photons from the annihilation of
slow positrons in the cavity. To cross-check this background, we estimate its level in the signal region 
by using available results from measurements of Ref.\cite{bader} and the proposal on the search for 
 $oPs \to invisible$  decay in vacuum of Ref.\cite{paolo} in a different way.
 In Fig. \ref{spect} the expected  distribution of energy deposition in the BGO calorimeter from the decays
 of $8\times 10^{12}$  $\mu^+$'s stopped in the target is shown. The spectrum represents the sum of 
 $\mu^+ \to all$ and $M\to all$ distributions . 
 The part of the spectrum above $\gtrsim $ 1 MeV is calculated from the Michel spectrum 
convoluted with the  ECAL (Gaussian) energy resolution.   
The peak around $\simeq 1$ MeV is from  the fraction  of decay positrons ($\simeq 10^{-4}$) with energy below of a few MeV  that are stopped in the cavity, i.e. either in the target or in the cavity walls,
 and annihilate into 2 or 3 photons. 
The  low energy tail below 1 MeV is described by a function  $f(E_{e^+}) =  f_1(E_{e^+}) +  f_2(E_{e^+})$, which  
is a  sum of two distributions of the annihilation energy in the ECAL normalized  to the same number of positrons annihilated in the cavity. The function $f_1(E_{e^+})$ is an experimentally measured distribution
  taken from the experiment on $Ps \to invisible$ \cite{bader} for  positrons annihilated in the SiO$_2$ target, 
  which did not take into account the annihilation photon absorption in the cavity walls. The function $f_2(E_{e^+})$
 is taken from the proposal \cite{paolo} and corresponds to the simulated energy deposition  in the ECAL
 minus energy absorbed in the cavity walls.  The sum function 
$f(E_{e^+})$ is then   extrapolated to zero energy resulting in a prediction of  about $8\pm2$ background events in the signal region for $8\times 10^{12}~\mu^+$'s stop in the target, which is somewhat smaller, but still in a reasonable agreement with the conservative number of about 18 events obtained from  from Table 1. The error of the above estimate  is defined by the uncertainty in the extrapolation procedure.

\begin{table}
\caption{\label{tab:table1} Expected contributions to the total level of
background from different background sources ( see text for details). }
\begin{ruledtabular}
\begin{tabular}{lr}
Source of background& Expected level\\
\hline
fake muon tag & $ \lesssim 10^{-13}$\\
inefficiency of of slow positrons detection \footnote{ The threshold for  energy deposited  in the ECAL from the decay $e^+$'s  annihilation is 100 keV.} & $ \lesssim 2\times  10^{-12}$\\
$e^+ +A\to invisible$  & $ \lesssim 10^{-13}$\\
ECAL hermiticity  & $\simeq 10^{-13}$\\
\hline 
Total ( conservatively)  &         $ \simeq  2.3  \times 10^{-12}$\\
\end{tabular}
\end{ruledtabular}
\end{table}

\subsection{Sensitivity of the proposed experiment}

 The significance of the $\mi$ decay  discovery  with such  a detector, 
% assuming $G_{M M'} \simeq 10^{-2}\cdot G_F$, 
 scales as  \cite{bk1,bk2} 
\begin{equation}
S=2\cdot(\sqrt{n_s + n_b}-\sqrt{n_b})
\label{sens}
\end{equation}
with 
\begin{equation}
n_s = n_\mu  \epsilon  f  Br(\mi) t
\end{equation} 
\label{sign}
and the branching ratio  of the muonium invisible decay  defined by
\begin{equation}
Br(M\to invisible) = \frac{n_s}{n_\mu \epsilon f t }
\label{br}
\end{equation} 
where  $n_s$ is the number of observed signal events (or the upper limit of the observed number of 
events), $n_b $ is the number of  background events, 
$n_\mu$ is the  muon beam intensity,  $t$ is the experiment running time,
 $\epsilon$ is the efficiency of the muonium production per incident muon, and factor $f$ corresponds either to the 
total number of decayed $M$ atoms ($f\simeq 1)$, or to the fraction of $M$ atoms that decay presumably in vacuum, 
not in the target ($f\simeq 0.033)$.

Before defining the expected sensitivity, let us first 
 discuss  several additional limitation factors.
The first one is related to the relatively long muon lifetime and the corresponding 
ECAL signal integration time. Indeed, 
 to get  the branching ratio $Br(\mi) \simeq 10^{-11}$ , the ECAL gate duration  
$\tau_g$, and hence the dead-time per trigger, has to be 
\begin{equation} 
\tau_g \gtrsim - \tau_\mu \times ln(Br(M \to invisible)) \simeq 60 ~ \mu s
\label{dur}
\end{equation}
 in order to avoid 
background from the muon decays outside the gate. 
The best sensitivity is expected  at integration gate $\tau \simeq 60~\mu$s; however, further, 
more complicated analysis compromising the level of this background and increasing of the pileup
noise might be necessary. The pileup energy, which  corresponds to energy deposited in the BGO ECAL by an additional 
undetected and uncorrelated particle, increases values of the ECAL pedestals. The amount of additional  energy in each BGO counter can be measured with the random trigger \cite{bader}. In the Ps experiment \cite{bader},
  for orthopositronium lifetime in the SiO$_2$ aerogel 
target of 132 ns the ECAL gate duration $\tau_{Ps}$ was chosen to be  $\simeq 2~\mu s$. 
This resulted in distribution of the sum of pedestals of all  
ECAL counters corresponding to the efficiency of ''zero'' signal detection as a function of  
the energy threshold. In order to keep the energy threshold as low as possible 
an algorithm to sum up the energy of all the ECAL crystals can be employed by exploiting the 
granularity of the calorimeter and fixing a zero energy threshold for each individual crystal. 
Taking into account the ECAL granularity, the effective ECAL energy  threshold can be  significantly 
reduced  from  80  keV, used to define the signal range for the $o-Ps \to invisible$ decay \cite{bader}, 
to about 20 keV having the overall signal efficiency above 95\% \cite{pcth}. In the proposed  experiment the
 longer gate will lead to 
 an increase of the pileup and pick-up electronic 
noise and hence to the overall
broadening of the  signal range,  approximately by a factor  
$\sqrt{\tau_g/\tau_{Ps}}\simeq 5$
and, hence to an increase of the effective energy threshold roughly up to
 $E_{th}\simeq 20 $ keV$\times 5 \simeq 100$ keV. 

Another limitation factor is related to the dead time of  Eq.(\ref{dur})
and, hence to  the maximally allowed muon 
counting rate, which  according to  Eq.(\ref{dur})  
 has to be  $ \lesssim 1/ \tau_g \simeq 10^{4}~ \mu^+ / s$ 
to avoid significant pileup effect.    
To minimize dead time, one could impose a time structure on the continuous beam by using 
a fast beam chopper operating in  a "muon on request" mode \cite{chopper}, and a first-level trigger rejecting events 
with the ECAL energy deposition greater than $E_{th}$ and, hence,   run the 
experiment at the rate $\simeq 1/\tau_\mu \simeq  5\times 10^5~\mu/s$. 
Assuming this rate, we anticipate   $8\times 10^{12}$ $\mu^+$ on target and 
production of about $6\times 10^{12}$ muonium atoms during 6 months of running time for the experiment. 
 Out of them, about $5.8\times 10^{12}$ $M$'s decay in the target, while about $2\times 10^{11}$ $M$'s leave the 
 target surface and decay in vacuum. For counting signal rate of $\simeq 10^{-11}$ per incident muon. assuming beam intensity of $\simeq 5\times 10^5~\mu^+$/s at $\simeq 90\%$ efficiency, it would require 1 week to accumulate 
 one signal event.  

In the background free experiment one could expect a sensitivity in the $M \to invisible$
 decay branching ratio of the order of 
\begin{equation}
Br(\mi) \lesssim 10^{-12},
\end{equation}
 assuming that in Eq.(\ref{br}) $n_s = 2.3$. For $M$'s that  decay in vacuum, the sensitivity is 
\begin{equation}
Br^{vac}(\mi) \lesssim 10^{-11}.
\label{vac}
\end{equation}
%It is difficult to evaluate the exact amount of beam time for a speculative experiment as this one. It is assumed, that 
%this experiment aims at  running almost 6 months  and  is  capable  to produce about $10^{11}$ muonium atoms in vacuum. 
 
In the presence of background and  in accordance with the SM prediction,  
the expected number of observed  events in the signal region $E \lesssim 100$ keV is 
\begin{equation}
N_M \simeq 50 \pm 7 ~ \rm{events}
\label{signal}
\end{equation}
out of which 18.4 events represent  conservatively estimated overall background from Table 1. 
Taking into account \eqref{sens}, one can see that the observation of the $\mi$ decay with about 5 $\sigma$ 
significance could be possible.

The statistical limit on the sensitivity of the proposed experiment to search for 
the decay $\mi$ due to transition into the hidden sector  is proportional to 
$G_{M M'}^2$ and is set by its value, see \eqref{mir}. Thus, to improve the sensitivity of \eqref{vac} 
 larger amount of muonium atoms decaying in vacuum  is required. Therefore, the improvement of the  efficiency for  thermal $M'$s production  is crucial for further searches. 
 
Note, that in the case 
of the signal observation,  to cross-check the result, one could replace the target with another one of the same
density, but not capable of  muonium producing, and run the experiment with suppressed $M$ decays, see e.g., Ref.\cite{opsmos}. In this case the distribution of the energy deposition in the ECAL, shown in Fig.\ref{spect}
would contain mainly events from the decays $\mu^+ \to all$ and the signal from the decays $\mi$
should disappear. In the case of observation  
of a higher than predicted $\mi$ decay rate, there is another important 
cross-check. Namely, as discussed in Sec. 3,  one could slightly modify the experimental conditions
without affecting the background , 
e.g. by  increasing  either the magnetic field in the cavity or 
the number of  muonium
 collisions with residual gas  molecules  by increasing the gas pressure \cite{paolo,dgt}.  These would  suppress
the muonium-mirror muonium oscillations, and the observed signal should vanish.

The performed analysis  
 gives an illustrative correct order of magnitude for the sensitivity of the 
proposed experiment. The simulations are performed without taking into account such effects as, 
e.g. pileup, and may be strengthened  by more accurate and  detailed Monte Carlo 
simulations of the concrete  experimental setup.\\

\section {Conclusion}

Due to its specific properties, muonium  is
 an important and interesting probe of the SM and  physics beyond the SM both from the 
 theoretical and  experimental view points.
 In the SM, the invisible decay $M\to \nut$ of muonium atoms into two neutrinos  is expected to be a very rare process  with the  branching fraction predicted to be $Br(M\to \nut) = 6.6 \times 10^{-12}$ with respect to the ordinary muon decay rate. This process has never been  experimentally tested.
Using the  reported experimental results on precision measurements of the positive muon lifetime by the MuLan 
Collaboration,  we set the  first limit   $Br(M \to invisible) < 5.7 \times 10^{-6}$, while  still leaving a big gap  of about six orders of magnitude between this bound and the predictions.  

To improve substantially the sensitivity, we  proposed to perform an experiment dedicated to 
the  search for the $\mi$ decay. The key point for the experiment is the presence of energy release from the annihilation of the low energy decay positrons in the detector.
A feasibility study of the experimental setup shows 
that the sensitivity  of the  search for this decay mode in branching fraction $Br(\mi)$
at the level of $10^{-12}$  could be achieved.  Thus, 
the SM prediction for the   $M\to invisible$  decay 
to exist at the level  of $ Br(\mi) \simeq 6.6 \times  10^{-12}$,  could be 
experimentally tested  for the first time.  
We point out that the  $\mi$ decay rate could be enhanced by non-SM contributions.  For instance, in the framework of the mirror matter model if the coupling strength between $M$ and  $M'$ is large enough, 
say $G_{MM'} \gtrsim 10^{-4} G_F$,  the decay $\mi$ could occur at a rate as high as the SM one.
If the proposed search results in a substantially higher branching 
fraction  than the SM predictions, say $Br(M \to invisible) \simeq 10^{-10}$ , 
this would unambiguously   indicate the presence of new physics. 
A result in agreement with the SM
prediction would provide a theoretically clean check of the pure leptonic bound state 
annihilation through charged current weak interactions, and provide constraints for further attempts beyond the SM.

The preliminary analysis  shows that the
quoted  sensitivity could be obtained with a detector  optimized for  
several of its properties. Namely, i) the primary beam and the entrance aperture 
size, ii) the efficiency of the muonium production in the target and in vacuum, iii)
the material composition and dimensions of the target and vacuum cavity, iv) the
efficiency of the veto counters S$_{1-4}$, and v) the pileup effect and zero-energy 
threshold in the ECAL are of importance.
  
 We believe our proposal, when paired with an existing  BGO calorimeter,  
   provides interesting  motivations for  the experiment on the  search for the $\mi$ decay to be performed in the near future.  
This low-energy experiment might be a sensitive  probe of new 
 physics that is complementary to  collider experiments. For example, it could also significantly improve
the recently obtained modest  bounds on the $\mu^+ \to invisible $  decay \cite{sngmu}, 
pushing it down to the region  $Br(\mu^+ \to invisible) \simeq 10^{-12}$.  A bound   in this  
region  will be  of  interest for several  extensions of the  Standard Model, see e.g., Ref. \cite{rub}.
The required high numbers of muonium atoms can be presently produced at PSI \cite{muconv}, or 
could  be available from high intensity  muon beams at future facilities such as the PRISM source at J-PARC 
\cite{prism}, the Project X at FNAL \cite{prx}, or  the neutrino factory \cite{mf}.\\
 
{\large \bf Acknowledgments}

We thank P. Crivelli, D. Gorbunov, F. Guber, A. Ivashkin, A. Rubbia, V. Samoylenko and D. Sillou for 
discussions. 
The help of  D. Sillou  and A. Korneev in calculations    is gratefully acknowledged.
This work was supported by RFBR grant N 10-02-00468a.

\end{document}